\documentstyle[sprocl]{article}

\def\Journal#1#2#3#4{{#1} {\bf #2}, #3 (#4)}

\def\CQG{\em Classical and Quantum Gravity}
\def\CMP{\em Commun. math. Phys.}
\def\JMP{\em J. Math. Phys.}

\def\e{{\rm e}}

\def\be{\begin{equation}}
\def\ee{\end{equation}}
\def\bea{\begin{eqnarray}}
\def\eea{\end{eqnarray}}

\begin{document}

\title{ON CYCLICALLY SYMMETRICAL SPACETIMES}

\author{A. BARNES}

\address{Computer Science, Aston University, 
Birmingham,\\ B4 7ET, UK\\E-mail: barnesa@aston.ac.uk} 

\maketitle\abstracts{In a recent paper Carot et 
al.~considered the definition of
cylindrical symmetry as a specialisation of the case of axial
symmetry.
One of their propositions states that if there is a second
Killing vector, which together with the one generating the
axial symmetry, forms the basis of a two-dimensional Lie algebra, then
the two Killing vectors must commute, thus generating an Abelian
group.
In this paper a similar result, valid under considerably weaker
assumptions, is derived: any two-dimensional Lie transformation group
which contains a one-dimensional subgroup whose orbits are circles,
must be Abelian.
The method used to prove this result is extended to apply to
three-dimensional Lie transformation groups. 
It is shown that the existence of a
one-dimensional subgroup with closed orbits restricts the Bianchi type
of the associated Lie algebra  to be  I, II, III,
VII$_{q=0}$, VIII or IX.    Some results on n-dimensional Lie groups are 
also derived and applied to show there are severe restrictions on
the structure of the allowed four-dimensional Lie transformation
groups compatible with cyclic symmetry.}

\section{Introduction}
Following Carter~\cite{bc} a spacetime $\cal M$ is said to have  
cyclical symmetry if and only if the metric is invariant under the
effective smooth action $SO(2) \times \cal M \rightarrow \cal M$ of the
one-parameter cyclic group $SO(2)$.  A cyclically symmetric spacetime
in which the set of fixed points of this isometry is not empty is said to be 
axially symmetric and the set of fixed points itself is referred to as
the  axis (of symmetry).  Mars and Senovilla~\cite{ms} proved a number of 
useful results on the structure of the axis.  Carot, 
Senovilla and Vera~\cite{csv} considered a definition of cylindrical 
symmetry based on the following proposition: if in an axial symmetric 
spacetime there is a second Killing vector which, with the Killing vector 
generating the axial symmetry, generates a two-dimensional isometry group 
then the two Killing vectors commute and the isometry group is Abelian.  A 
similar result for stationary axistymmetric spacetimes was proved by 
Carter~\cite{bc}.

The proofs of all the above mentioned results rely heavily on the existence 
of an axis and although the assumption of the existence of an axis is 
reasonable in many circumstances, there are numerous situations where an axis 
in a cyclically symmetric manifold may not exist.  The 'axis' may be 
singular due to line sources and so not part of the manifold proper or the 
topology of the manifold may be such that no axis exists as is the case for 
the standard two-dimensional torus embedded in three-dimensional Euclidean 
space.  In the next section the condition for the existence of an axis 
will be discarded and the following result will be proved: any
two-dimensional Lie transformation group which acts on an
$n$-dimensional manifold $\cal M$ and which contains a one-dimensional
subgroup acting  cyclically on $\cal M$ must be Abelian. 
In subsequent sections 
three and higher dimensional Lie transformation groups will be considered 
and their structure shown to be severely restricted by the existence of a 
one-dimensional subgroup with circular orbits.

\section{Cyclically Symmetric Manifolds Admitting a $G_2$}
Suppose $X_0$ is the Killing vector field associated with the cyclic 
isometry acting on $\cal M$ and let $\cal N$ be the open submanifold of 
$\cal M$ on which $X_0 \ne 0$.  The orbit of each point of $\cal N$ under 
the cyclic symmetry is a circle.  Let $\phi$ be a circular coordinate 
running from $0$ to $2\pi$ which parameterises elements of $SO(2)$ in the 
normal way.  Then we can introduce a coordinate system $x^i$ with 
$i=1\ldots n$ and $x^1=\phi$ adapted to $X_0$ such that $X_0 = 
\partial_\phi$.

Suppose that the isometry group of $\cal M$ admits a two-dimensional 
subgroup $G_2$ containing the cyclic symmetry and let $X_0$ and $X_1$ be a 
basis of the Lie algebra of $G_2$. In an adapted coordinate system the 
commutator relation of the Lie algebra of $G_2$,
\be
[X_0\ X_1] = aX_0 + bX_2
\label{eq:com2}
\ee
where $a$ and $b$ are constants, reduces to
$$
{\partial X_1^\mu \over \partial \phi} = b X_1^\mu \qquad 
{\partial X_1^1 \over \partial \phi} = a + b X_1^1
$$
where Greek indices range over the values $2\ldots n$.
An elementary integration gives
$$
\begin{array}{rclrclr}
X_1^\mu & = &B^\mu(x^\nu)\e^{b\phi}\qquad & X_1^1 & =
 &A(x^\nu)\e^{b\phi} + a/b    \qquad &{\rm for\ } b \ne 0\\
X_1^\mu & = & B^\mu(x^\nu) & X_1^1 & = & A(x^\nu) + a\phi & 
{\rm for\ } b =  0
\end{array}
$$
where $A$ and $B^\mu$ are arbitrary functions of integration.
If $X_1$ is to be single-valued on $\cal N$, then these solutions must
be periodic in $\phi$ with period $2\pi$.  This can only occur if
$a=b=0$ and so from Eq.~(\ref{eq:com2}) $G_2$ must be Abelian.

The dimensionality of the manifold, the existence of a metric  and the
fact that the transformation group is an isometry group are not used
in the proof. Hence we have shown that any
two-dimensional Lie transformation group which acts on an
$n$-dimensional manifold $\cal M$ and which contains a one-dimensional
subgroup with cicular orbits, must be Abelian. Thus {\it a
fortiori\/} the result holds when the $G_2$ is group of motions, 
conformal motions, affine collineations or projective collineations.
This remarkably simple and general result has appeared in the
literature previously (for example Bi\v c\'ak \& Schmidt~\cite{bs}) but is
perhaps not widely known.  

\section{Cyclically Symmetric Manifolds Admitting a $G_3$}
Suppose that $\cal M$ admits a three-dimensional Lie
transformation group $G_3$ containing a one-dimensional subgroup acting
cyclically on $\cal M$ generated by the vector field $X_0$.
Let $X_1$ and $X_2$ be vector fields on $\cal M$ which, with $X_0$,
form a basis of the Lie algebra of $G_3$.  Now either this Lie algebra
admits a two-dimensional subalgebra containing $X_0$ or there is no
such subalgebra.

In the former case this subalgebra is Abelian by the result proved in
the previous section.  We may assume, without loss of generality, that
$X_0$ and $X_1$ form a basis of this subalgebra and consequently the
commutation relations involving $X_0$ can be written in the form
$$ [ X_0\ X_1] = 0 \qquad [X_0\ X_2]  = a X_0 + b X_1 + c X_2 $$
where $a$,$b$ and $c$  are constants.  In a coordinate
system adapted to $X_0$ in which $X_0 = \partial_\phi$ these equations become
$$
{\partial X_1^\i \over \partial \phi} = 0 \qquad 
{\partial X_2^i \over \partial \phi} = a\delta^i_1 + b X_1^i + c X_2^i
$$
On integrating these equations and using the fact that $X_1$ and $X_2$
must be periodic in $\phi$ with period $2\pi$, we may deduce that
$a=b=c=0$.  Thus $X_0$ commutes with both $X_1$ and $X_2$.
The remaining basis freedom preserving $X_0$ is
$$
\tilde X_1 = \alpha X_1 + \beta X_2 + \lambda X_0 \qquad
\tilde X_2 = \gamma X_1 + \delta X_2 + \mu X_0 
$$
subject to the condition $\alpha\delta - \beta\gamma \ne 0$.
Using this basis freedom we may reduce the commutators of the Lie
algebra of $G_3$ to one of the following forms 
$$
\begin{array}{rclrclrcll}
\lbrack X_0\ X_1] & = & 0 \qquad & [X_0\ X_2] & = & 0 
\qquad & [X_1\ X_2] & = & 0 \qquad & {\rm Bianchi\ type\ I}\\
\lbrack X_0\ X_1] & = &0 \qquad & [X_0\ X_2] & = &0 
\qquad & [X_1\ X_2] & = & X_0 \qquad &{\rm Bianchi\ type\ II}\\
\lbrack X_0\ X_1] & = & 0  \qquad & [X_0\ X_2] & = & 0
\qquad &[X_1\ X_2] & = & X_2  \qquad & {\rm Bianchi\ type\ III}
\end{array}
$$
These are the canonical forms for the commutators of Bianchi types I, II and
III algebras given by Petrov~\cite{azp} (apart from renumbering of the basis
vectors for type III).

If the Lie algebra of $G_3$ has no two-dimensional subalgebra
containing $X_0$, we may always choose basis vectors $X_1$ and $X_2$
such that the commutation relations involving $X_0$ become
$$ [ X_0\ X_1] = X_2 \qquad [X_0\ X_2]  = a X_0 + b X_1 + c X_2 $$
where $a$, $b$ and $c$ are constants.  In terms of a coordinate system
adapted to $X_0$ in which $X_0 = \partial_\phi$ these become
$$
{\partial X_1^\i \over \partial \phi} = X_2^i \qquad 
{\partial X_2^i \over \partial \phi} = a\delta^i_1 + b X_1^i + c X_2^i
$$
A straightforward integration of these equations reveals that, for
solutions periodic in $\phi$ with period $2\pi$, we must have $c=0$
and $b=-n^2$ for some positive integer $n$.  Then by a redefinition of
the basis vector $\tilde X_1 = n X_1 - a/n X_0$,
we can set $a=0$. Hence the commutation relations become
$$[X_0\ X_1] = n X_2 \qquad [X_0\ X_2] = -n X_1 
\qquad [X_1\ X_2] = d X_0 + e X_1 + fX_2$$
where $d$, $e$ and $f$ are constants. The Jacobi identity 
$$[X_0\ [X_1\ X_2]] + [X_1\ [X_2\ X_0]] + [X_2\ [X_0\ X_1]] = 0$$
implies that
$$[X_0\ \ (d X_0 +e X_1 +f X_2)] = n (e X_2 -  f X_1) = 0$$
Thus $e = f = 0$.  Three algebraic distinct types arise: namely Bianchi types
VII${_q=0}$, VIII or IX depending on whether $d=0,\ <0,\ > 0$
respectively. The commutation relations may be written in one of the
following  forms
$$
\begin{array}{rclrclrcll}
\lbrack X_0\ X_1] & =& n X_2  \quad & [X_0\ X_2] & = & -n X_1 
\quad & [X_1\ X_2] & = & 0 \quad & {\rm Bianchi\ type\ VII}_{q=0}\\
\lbrack X_0\ X_1] & = &n X_2 \quad & [X_0\ X_2] & = &-n X_1 
\quad & [X_1\ X_2] & = & -X_0 \quad &{\rm Bianchi\ type\ VIII}\\
\lbrack X_0\ X_1] & = & n X_2  \quad & [X_0\ X_2] & = & -n X_1
\quad &[X_1\ X_2] & = & X_0  \quad & {\rm Bianchi\ type\ IX}
\end{array}
$$
where, in the last two types we have set $d = \mp 1$ by the rescaling
$$\tilde X_1 = 1/\sqrt{|d|} X_1\qquad \tilde X_2 = 1/\sqrt{|d|} X_2$$
These commutators are closely related to the canonical forms
of Bianchi types VII$_{q=0}$, VIII and IX given by
Petrov~\cite{azp}. To get the canonical forms we would need to scale
$X_0$ to set  $n=1$.  However this cannot be done whilst preserving
both the equation $X_0 = \partial_\phi$ and the $2\pi$ periodicity of
the coordinate $\phi$.

Only six of the nine Bianchi types can occur; Bianchi types IV, V and
VI are excluded.  Note also that 
canonical forms of the algebras of Bianchi types VI and VII depend on
an arbitrary real parameter $q$ and so each contain an infinite number
of algebraically distinct cases;
those in type VI are excluded completely and of those in type VII only
a single case, $q=0$, can occur. Moreover in all the Bianchi types that are
permitted, the cyclic vector $X_0$ is aligned with a vector of a basis in
which the commutation relations take their canonical form.

\section{Cyclically symmetric manifolds admitting a $G_{m+1}$}
Suppose now that $\cal M$ admits an $(m+1)$-dimensional Lie
transformation group $G_{m+1}$ containing a one-dimensional subgroup acting
cyclically on $\cal M$ generated by the vector field $X_0$.
Let $X_a$ be vector fields on $\cal M$ which, with $X_0$,
form a basis of the Lie algebra of $G_{m+1}$.  Here and below indices
$a$, $b$ and $c$ take values in the range $1\ldots m$.
The commutators involving $X_0$ may be written in the form
\be
[X_0\ X_a] = C^b_a X_b + D_a X_0
\label{eq:gcr}
\ee
where $C^b_a$ and $D_a$ are constants.   

If we introduce new basis vectors $\tilde X_a$ given by
$\tilde X_a = P^b_a X_b$, the structure constants transform as follows
$$\tilde {\bf C} = {\bf P} {\bf C} {\bf P}^{-1}
\qquad \qquad \tilde {\bf D} = {\bf P}{\bf D}$$
where for simplicity we have used standard matrix notation.
Using these transformations we can reduce ${\bf C}$ to Jordan
normal form.  In what follows we will work in a basis in which the structure
 constants $C^b_a$ are in Jordan normal  form but, for typographic
 simplicity tildes will be omitted. 

In terms of a coordinate system  adapted to $X_0$ in which  $X_0 =
 \partial_\phi$ the commutation relations in Eq.~(\ref{eq:gcr}) become
\be
{\partial X_a^\mu \over \partial \phi} = C^b_a X_b^\mu \qquad 
{\partial X_a^1 \over \partial \phi} = C^b_a X_b^1 + D_a
\label{eq:diffcr}
\ee
where Greek indices range over $2\ldots n$.  If the solutions of these
equations are to be periodic in $\phi$ with period $2\pi$, then the eigenvalues
$\lambda$ of ${\bf C}$ must either be zero or of the form $\lambda =
\pm i n$ where n is a positive integer.  Moreover  all of
the Jordan blocks must be simple or equivalently the minimal polynomial
of ${\bf C}$ must have no repeated factors.

Suppose without loss of generality that ${\bf C}$ has $p$ 
($0 \le  2p \le m$) eigenvalues of the form $i n_j$ ($n_j$ positive 
 integers and $1 \le j \le p$) with  corresponding complex eigenvectors 
$Z_j \equiv  X_{2j} + i X_{2j-1}$ plus $m-2p$ zero eigenvalues with corresponding
real eigenvectors $X_k$  ($2p+1 \le k \le m$).  Choosing these $m$
 (real) vectors $X_a$ as the basis vectors, the commutators become
\be
\begin{array}{lcll}
\lbrack X_0\ X_{2j-1}] &=& n_j X_{2j}\qquad & {\rm for\ }1 \le j \le p\\
\lbrack X_0\ X_{2j}] &=& -n_j X_{2j-1}\qquad & {\rm and\ }0 \le 2p \le m\\
\lbrack X_0\ X_k] &=& 0 \qquad & {\rm for\ }2p+1 \le k \le  m
\end{array}
\label{eq:ccr}
\ee
In the above commutators the structure constants $D_a$ that appeared
in Eq.~(\ref{eq:gcr}) have been set to zero. This is valid since, for 
$2p+1 \le k \le m$, the vanishing of $D_k$ is a consequence of the
periodicity of the solution of the second of Eqs.~({\ref{eq:diffcr})
which reduces to
$${\partial X_k^1 \over \partial \phi} = D_k$$
For $1 \le j \le p$, $D_{2j-1}$ and $D_{2j}$ can be set to zero
by a transformation of the basis vectors of the form
$$\tilde X_{2j-1} = X_{2j-1} - D_{2j}/n_j X_0 \qquad 
\tilde X_{2j} = X_{2j} + D_{2j-1}/n_j X_0 $$

Thus even in the general case the existence of a one-dimensional group
acting cyclically on the $\cal M$ imposes quite strong restrictions on
the allowed form of the commutation relations involving $X_0$. 
Also the results of sections 2 and 3 can be seen to be special cases of the
general result just proved.  

For the $G_4$ case ($m=3$) two classes of algebra arise with commutators
$$
\begin{array}{lcllcllcl}
\lbrack X_0\ X_1] &=& 0 \qquad &[X_0\ X_2] &=& 0 
\qquad &[ X_0\ X_3] &=& 0 \\
\lbrack X_0\ X_1] &=& n X_2 \qquad &[X_0\ X_2] &=& -n X_1 
\qquad &[ X_0\ X_3] &=& 0 
\end{array}
$$
corresponding to the cases $p=0$ and $p=1$ in Eq.~(\ref{eq:ccr}) respectively. 

The Jacobi identities further restrict the structure constants
appearing in the remaining commutators. In fact it 
is possible to enumerate completely the four-dimensional algebras of $G_4$
groups compatible with cyclic symmetry and to relate them to the eight
types listed by Kruchkovich~\cite{gik} and Petrov~\cite{azp} in their complete
classification of {\it all\/} 
four-dimensional Lie algebras. Examples of all eight of the types can
occur, but in many cases only a zero-parameter or 
one-parameter subset of a two-parameter Kruchkovich-Petrov type is
permitted and some subclasses of the Kruchkovich-Petrov types are
excluded completely.
Furthermore the vector $X_0$ generating the cyclic subgroup is always
nicely aligned with the canonical bases used by Kruchkovich and Petrov.  A
complete account of the results for $G_4$ will appear elsewhere.

\end{document}